\def\MPL #1 #2 #3 {Mod.~Phys.~Lett.~{\bf#1},\  #2 (#3)}
\def\NPB #1 #2 #3 {Nucl.~Phys.~{\bf#1},\  #2 (#3)}
\def\PLB #1 #2 #3 {Phys.~Lett.~{\bf#1},\  #2 (#3)}
\def\PR #1 #2 #3 {Phys.~Rep.~{\bf#1},\ #2 (#3)}
\def\PRD #1 #2 #3 {Phys.~Rev.~{\bf#1},\  #2 (#3)}
\def\PRL #1 #2 #3 {Phys.~Rev.~Lett.~{\bf#1},\  #2 (#3)}
\def\RMP #1 #2 #3 {Rev.~Mod.~Phys.~{\bf#1},\  #2 (#3)}
\def\ZP #1 #2 #3 {Z.~Phys.~{\bf#1},\  #2 (#3)}
\def\IJMP #1 #2 #3 {Int.~J.~Mod.~Phys.~{\bf#1},\  #2 (#3)}
\def\br{BR}
\def\anti{\overline}

\def\zstar{Z^{\star}}

\def\rts{\sqrt s}

\def\h{h}
\def\mh{m_{\h}}

\def\gamhtot{\Gamma_{\h}^{\rm tot}}

\def\eg{{\it e.g.}}

\def\epem{e^+e^-}

\def\lsim{\mathrel{\raise.3ex\hbox{$<$\kern-.75em\lower1ex\hbox{$\sim$}}}}
\def\gsim{\mathrel{\raise.3ex\hbox{$>$\kern-.75em\lower1ex\hbox{$\sim$}}}}
\def\@versim#1#2{\vcenter{\offinterlineskip
        \ialign{$\m@th#1\hfil##\hfil$\crcr#2\crcr\sim\crcr } }}
\def\zstar{Z^\star}

\def\ie{{\it i.e.}}

\def\gam{\gamma}

\def\anti{\overline}

\def\fbi{~{\rm fb}^{-1}}

\def\gev{\,{\rm GeV}}

\def\hsm{h_{SM}}
\def\mhsm{m_{\hsm}}
\def\hl{h^0}

\def\mz{m_Z}

\def\wp{W^+}
\def\wm{W^-}

\def\h{h}
\def\mh{m_{\h}}

\def\ptmin{p_T^{\rm min}}
\def\dmgamgam{\Delta\mgamgam}
\def\mgamgam{m_{\gam\gam}}

\documentclass[smus]{snow2e}
\input psfig.sty
\begin{document}

\title{
{\normalsize
             \hspace*{\fill} UCD-96-34 \\
             \hspace*{\fill} October, 1996 \\
             }
Measuring the Higgs to {\boldmath $\gam\gam$} Branching Ratio
at the Next Linear {\boldmath $\epem$} Collider.
\thanks{
To appear in ``Proceedings of the 1996 DPF/DPB Summer Study
on New Directions for High Energy Physics''.
Work supported in part by the Department of Energy
and by the Davis Institute for High Energy Physics.}}

\author{J.~F. Gunion and  P.~C. Martin \\
{\it Davis Institute for High Energy Physics, 
University of California, Davis CA, 95616} 
}

\maketitle

%% Get rid of page numbering
\thispagestyle{empty}\pagestyle{plain}

\begin{abstract} 
We examine the prospects for measuring the $\gam\gam$
branching ratio of a Standard-Model-like Higgs boson ($\h$)
at the Next Linear $\epem$ Collider when the Higgs boson is produced 
via $\wp\wm$--fusion: $e^+e^-\to\nu_e \bar\nu_e \h$. In 
particular, we study the accuracy
of such a measurement and the statistical significance
of the associated signal as a function
of the electromagnetic calorimeter resolution and
the Higgs boson mass.
We compare results for the $\wp\wm$--fusion production/measurement mode with
the results obtained for the $\epem\to \zstar\to Z\h$ 
production/measurement mode in a parallel earlier study.
\end{abstract}

\section{Introduction}

Discovery and study of Higgs boson(s) will be of primary
importance at a Next Linear $\epem$ Collider (NLC). After discovery
of a Higgs,
the goal will be to determine as precisely as possible---independent
of any model---its fundamental couplings and total width.
Our concern is with a light Standard Model (SM)
like Higgs boson which has a width too small for 
direct observation \cite{hhg}. For such a Higgs boson, it will be necessary 
to determine $\br(\h\to\gam\gam)$ in order to determine
its total width and coupling constants.
The procedure for ascertaining the Higgs total width and
its $b\anti b$ partial width is outlined below.
(Estimated errors given are summarized in Ref.~\cite{dpfreport}.)
\begin{itemize}
\item Measure $\sigma(Z\h)$ (in the missing mass mode)
and $\sigma(Z\h)\br(\h\to b\anti b)$ and compute:
\begin{equation}
\br(\h\to b\anti b)={[\sigma(Z\h)\br(\h\to b\anti b)]\over \sigma(Z\h)}\;;
\label{stepi}
\end{equation}
the error in $\br(\h\to b\anti b)$ so obtained is estimated
at $\pm 8\%$ to $\pm 10\%$.
\item Measure at the associated $\gam\gam$ collider facility the
rate for $\gam\gam\to\h\to b\anti b$ (accuracy $\pm5\%$) which is
proportional to $\Gamma(\h\to\gam\gam)\br(\h\to b\anti b)$ 
and compute (accuracy $\pm 11\%$ to $\pm 13\%$):
\begin{equation}
\Gamma(\h\to \gam\gam)={[\Gamma(\h\to\gam\gam)\br(\h\to b \anti b)]\over
\br(\h\to b\anti b)}\;.
\label{stepii}
\end{equation}
\item Measure in $\epem\to \nu_e\anti\nu_e \h$ ($\wp\wm$--fusion)
the event rates for $\h\to\gam\gam$ and $\h\to b \anti b$. Then compute:
\begin{equation}
\br(\h\to \gam\gam)={\br(\h\to b\anti b)[\sigma(\nu_e \bar\nu_e \h)
\br(\h\to\gam\gam )]\over
[\sigma(\nu_e \bar\nu_e \h)
\br(\h\to b\anti b)]}\;.
\label{stepiii}
\end{equation}
\item Finally, compute:
\begin{equation}
\gamhtot= {\Gamma(\h\to\gam\gam)\over \br(\h\to\gam\gam)}\,;~~
\Gamma(\h\to b\anti b)=\gamhtot \br(\h\to b\anti b) \;.
\label{stepiv}
\end{equation}
\end{itemize}

The above technique determines both 
$\gamhtot$ and $\Gamma(\h\to b \anti b)$ in a model-independent way.
This is desirable since
knowledge of these fundamental Higgs properties is likely
to be far more revealing than a simple measurement of
$\br(\h\to b\anti b)$ alone.
For example, in the minimal supersymmetric
model (MSSM) parameters can be chosen such that the light Higgs,
$\hl$, has total width and $b\anti b$ partial width 
that are both significantly different
from the SM prediction, whereas the $b\anti b$ branching ratio is not.
This occurs because the numerator and denominator,
$\Gamma(\h\to b\anti b)$ and $\gamhtot$, respectively, 
differ by similar amounts from the SM predictions, so that
the ratio of the two changes only slightly. In general,
interpretation of any branching ratio is ambiguous. We must
be able to convert the measured branching ratios to the partial
widths that are directly related to fundamental couplings.
This is only possible if we can determine $\gamhtot$ in a model-independent
way.

Estimating the error in the determination of $\br(\h\to\gam\gam)$
and how it propagates into errors in the determination of the total width
and thence partial widths is very crucial.
This is because the deviation of 
$\br(\h\to\gam\gam)$ and the partial widths
of a SM-like Higgs of an extended model from the predictions for the minimal
SM Higgs boson may be small (as typical, for example,
in the case of the $\hl$ of the MSSM when the pseudoscalar
Higgs boson of the MSSM is heavy). It turns out that the dominant
error in the partial width determinations 
will be that from the determination of $\br(\h\to\gam\gam)$. 
Thus, it is vital that we determine
the optimal procedures for minimizing the error in the latter.

Of course, deviations of $\br(\h\to\gam\gam)$ itself from
SM expectations could also be very revealing. In particular,
by virtue of the fact that the coupling $\h\to\gam\gam$
arises from charged loops, large deviations from SM predictions
due to new particles (\eg\ fourth generation, supersymmetry {\it etc.})
are possible.
Regardless of the size of the deviations from SM predictions,
determining $\br(\h\to\gam\gam)$ at the NLC will be vital to
understanding the nature of the Higgs boson and will
provide an important probe of new physics that may lie beyond the SM.

\section{SM Signal and Background } 

In this report we examine expectations in the case of
the Standard Model Higgs boson, $\hsm$. We focus on
the mass range $50\gev\lsim \mhsm\lsim 150\gev$ for which
$\br(\hsm\to\gam\gam)$ is large enough to be potentially measurable.
For the $\epem\to \nu_e\anti\nu_e\hsm$ process
that we are considering, the best rate is obtained by running
the $\epem$ collider at the maximum possible energy. We
adopt the canonical NLC benchmark energy of
$\rts  = 500 \gev$. 

Exact matrix elements are used for all
calculations. For completeness, when calculating the signal ($S$)
in the $X\hsm$ final state (where $X$ is invisible),
we include both production processes,
\begin{equation} 
e^+e^-\to\wp\wm\nu_e\bar\nu_e\to \nu_e \bar\nu_e \hsm\,, 
\end{equation}
\begin{equation}
e^+e^-\to \zstar \to Z\hsm \,,
\end{equation}
with the subsequent decays:
\begin{equation}
\hsm\to\gam\gam \,~~{\rm and}~~ Z \to \nu_i \bar \nu_i
~~ ~~ (i=e,\mu,\tau).
\end{equation}
When calculating the background ($B$) we include all processes
contributing to
\begin{equation}
e^+e^-\to \nu_i \bar \nu_i \gam\gam.
\end{equation}
In our parallel study of the $Z\hsm$ production/measurement mode, 
visible as well as invisible $Z$
decays were included in both signal and background.  
\footnote{In this case, the $Z$-pole
contributions to signal and background can be isolated 
for both visible and invisible final $Z$ decays by
requiring that the reconstructed `$Z$' mass computed from
the observed four-momenta of the photons and the incoming
$e^+$ and $e^-$ be near $\mz$.}

\section{Cuts and Calorimetry Considerations}

We compute both the signal and background rates for a small interval
of the two-photon invariant mass, $\dmgamgam(\mhsm)$, centered around 
$\mhsm$.\footnote{The Higgs mass will be measured very precisely
using the missing-mass technique in the $Z\hsm$ mode.} 
The $\dmgamgam$ interval 
will depend upon the resolution of the electromagnetic
calorimeter (as we shall shortly discuss) and is adjusted
in conjunction with other kinematic cuts so that the statistical
error in measuring $\sigma(\epem\to \hsm
X)\br(\hsm\to\gam\gam)$, $\sqrt{S+B}/S$, is minimized.
After exploring a wide variety of possible cuts, we found that
the smallest error could be achieved using the following:
\begin{equation} 
|y_{\gam_1}|\leq 2.0 \,,\quad |y_{\gam_2}|\leq 2.0\,,
\end{equation}
\begin{equation} 
 p_{T}^{\gam_{1,2}}\geq p_{T}^{\gam_{1,2}~{\rm min}}(\mhsm)\,,\quad
p_{T}^{\gam_1}+p_{T}^{\gam_2} \geq \ptmin(\mhsm)\,,
\label{cutsi}
\end{equation}
\begin{equation}
M_{missing}=\sqrt{(p_{e^+}+p_{e^-}-p_{\gam_1}-p_{\gam_2})^2} \geq 130\gev\,,
\end{equation}
\begin{equation}
p_{T}^{\rm vis}=\sqrt{(p_x^{\gam_1}+p_x^{\gam_2})^2+(p_y^{\gam_1}+p_y^{\gam_2})^2}
\geq 10 \gev\,,
\end{equation}
where $p_{T}^{\gam_{1,2}}$  are the magnitudes of the transverse momenta
of the two photons in the $\epem$ center-of-mass
(by convention,  $E_{\gam_1}\geq E_{\gam_2}$). 
The $M_{missing}$ cut effectively
removes contributions from $e^+e^-\to \zstar \to Z\hsm$ 
and the associated $\epem\to Z\gam\gam\to \nu\anti\nu \gam\gam$
backgrounds.  This is desirable because at $\rts=500\gev$
the $S/B$ ratio for these $Z$-pole-mediated processes is much smaller
than that for the $\wp\wm$--fusion signal contribution and non-$Z$-pole 
backgrounds.
Finally, the $p_{T}^{\rm vis}$ cut is used to eliminate contributions from
events such as $e^+e^- \to e^+e^- \gam\gam $
where the $e^+$ and $e^-$ are lost down the beam pipe
leaving the signature of $\gam\gam$ plus
missing energy \cite{ptvs}.

Four different electromagnetic calorimeter resolutions
are considered:
\begin{description}
\item{ I:} resolution like that of the CMS lead tungstate crystal \cite{CMS}
with $\Delta E/E=2\%/\sqrt E\oplus 0.5\%\oplus20\%/E$;
\item{ II:} resolution of $\Delta E/E=10\%/\sqrt E\oplus 1\%$; 
\item{ III:} resolution of $\Delta E/E=12\%/\sqrt E\oplus 0.5\%$; and
\item{ IV:} resolution of $\Delta E/E=15\%/\sqrt E\oplus 1\%$.
\end{description}
Cases II and III are at the `optimistic' end of current NLC detector
designs \cite{NLCdetector}. 
Case IV is the current design specification for the 
JLC-1 detector \cite{JLCdetector}. For each resolution case, 
we have searched for the 
$p_T^{\gam_1~{\rm min}}$, $p_{T}^{\gam_2~{\rm min}}$, $p_T^{\rm min}$
and $\dmgamgam$ values which
minimize the error, $\sqrt{S+B}/S$, at a given Higgs boson mass.
In Table~\ref{tablei}, we give these values as a function
of Higgs mass $\mh$. 
Listed in Table~\ref{tableii} are the signal and background rates
for the $\hsm$ computed for these optimal choices. 
We assume that $\mhsm$ will be known within $\Delta\mhsm\ll
\dmgamgam$ and that the backgrounds can be accurately
determined using data away from $\mhsm$. 
All the results are for four years of
running at $L=50\fbi$ yearly integrated luminosity, \ie\ a total
of $L=200\fbi$.

\section{Results and Discussion}

We present the statistical errors for measuring
$\sigma\br(\hsm\to\gam\gam)$ in the $\wp\wm$--fusion measurement mode
and compare with the results from our earlier, similar
study of the $Z\hsm$ production measurement mode \cite{zhmode}. 
We note that in the $Z\hsm$ case the optimal results
to be reviewed are only obtained by tuning the machine energy
close to the value which maximizes the $Z\hsm$ cross section for the
given value of $\mhsm$ and accumulating $L=200\fbi$ at that energy. 
(The exact $\rts$ values employed for the $Z\hsm$ measurement
mode and the associated cuts, the
nature of which differ somewhat from the ones presented here
for the fusion mode, are detailed in Ref.~\cite{zhmode}.)
Since the optimal $\rts$ for the $Z\hsm$ mode is 
always substantially less than $500 \gev$, 
the devotion of so much luminosity to this single
$\rts$ value will only take place once the $\hsm$ has already
been discovered at the LHC or while running the NLC 
at $\rts=500\gev$.  If the NLC is first operated at $\rts=500\gev$,
either because a Higgs boson has not been detected previously
or because other physics (\eg\ production of supersymmetric particles)
is deemed more important, data for measuring $\sigma\br(\hsm\to\gam\gam)$
using the $\wp\wm$--fusion measurement mode will be accumulated.
We will see that both the detector resolution and the actual
value of $\mhsm$ will enter into the decision regarding whether
or not to devote luminosity to the $Z\hsm$ measurement mode
at a lower $\rts$.

\begin{figure}[htb]
\leavevmode
\begin{center}
\centerline{\psfig{file=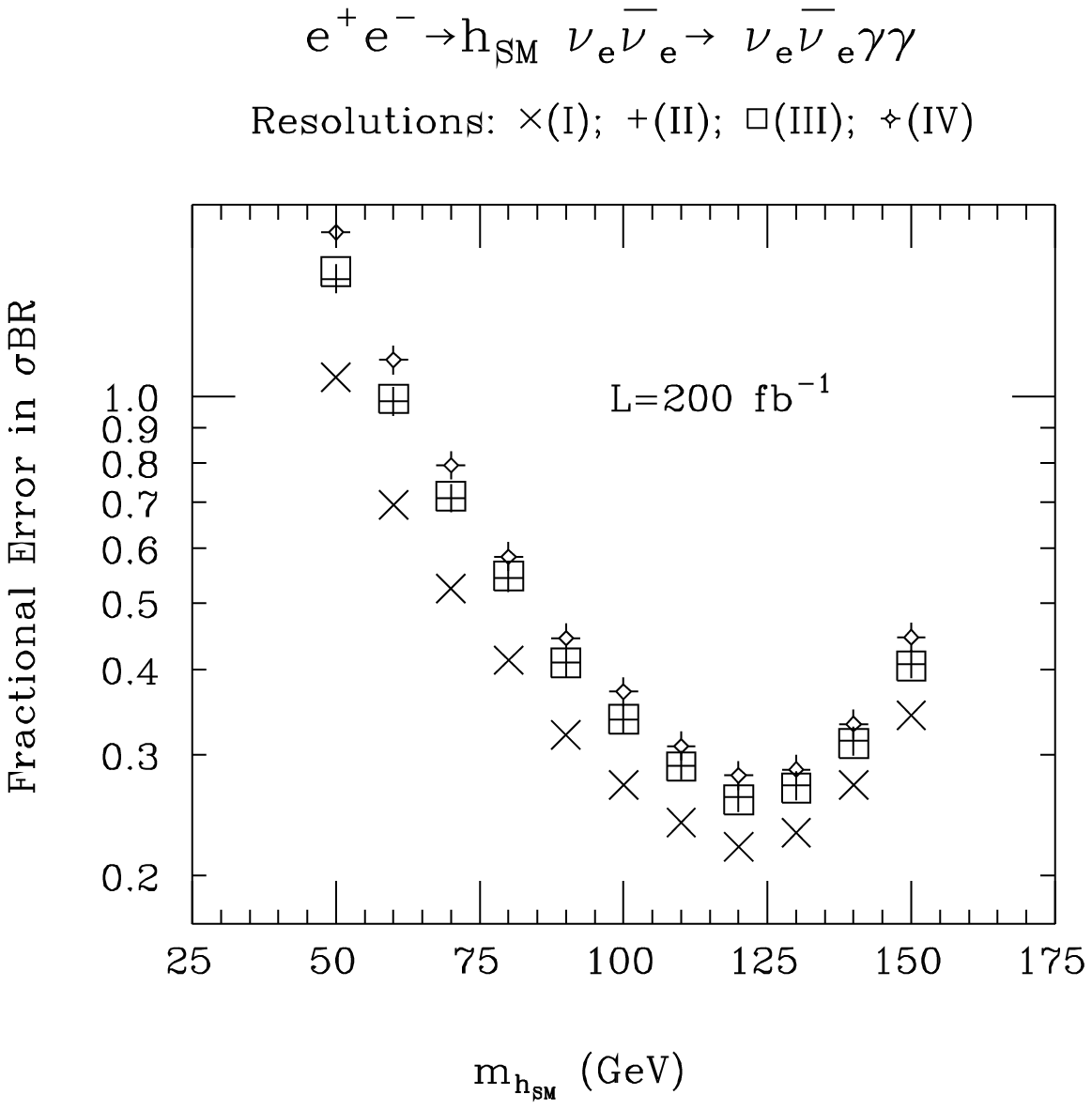,width=3.5in}}
%\bigskip
\end{center}
%\resizebox{!}{11cm}{%
%\includegraphics{3144A116.eps}}
%\end{center}
\caption{The fractional error in the measurement of
$\sigma(\nu_e \bar\nu_e \hsm)q
\br(\hsm\to\gam\gam)$ 
as a function of $\mhsm$.}
\label{fig1}
\end{figure}

\begin{figure}[htb]
\leavevmode
\begin{center}
\centerline{\psfig{file=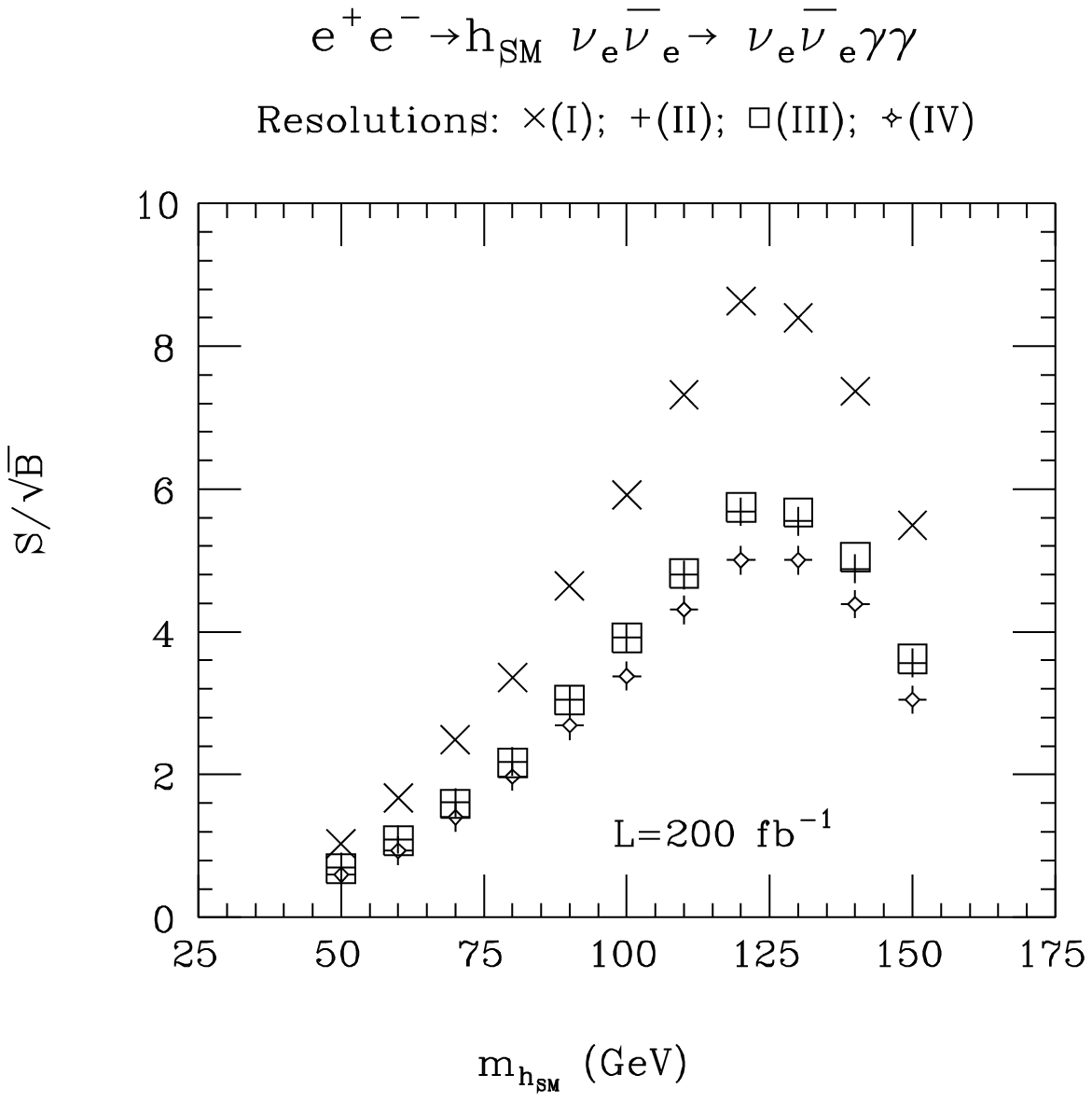,width=3.5in}}
%\bigskip
\end{center}
%\resizebox{!}{11cm}{%
%\includegraphics{3144A116.eps}}
%\end{center}
\caption{Results for $S/\protect\sqrt B$ in the $\wp\wm$--fusion
production mode at $L=200\fbi$ as a function of $\mhsm$.}
\label{fig2}
\end{figure}

\begin{figure}[htb]
\leavevmode
\begin{center}
\centerline{\psfig{file=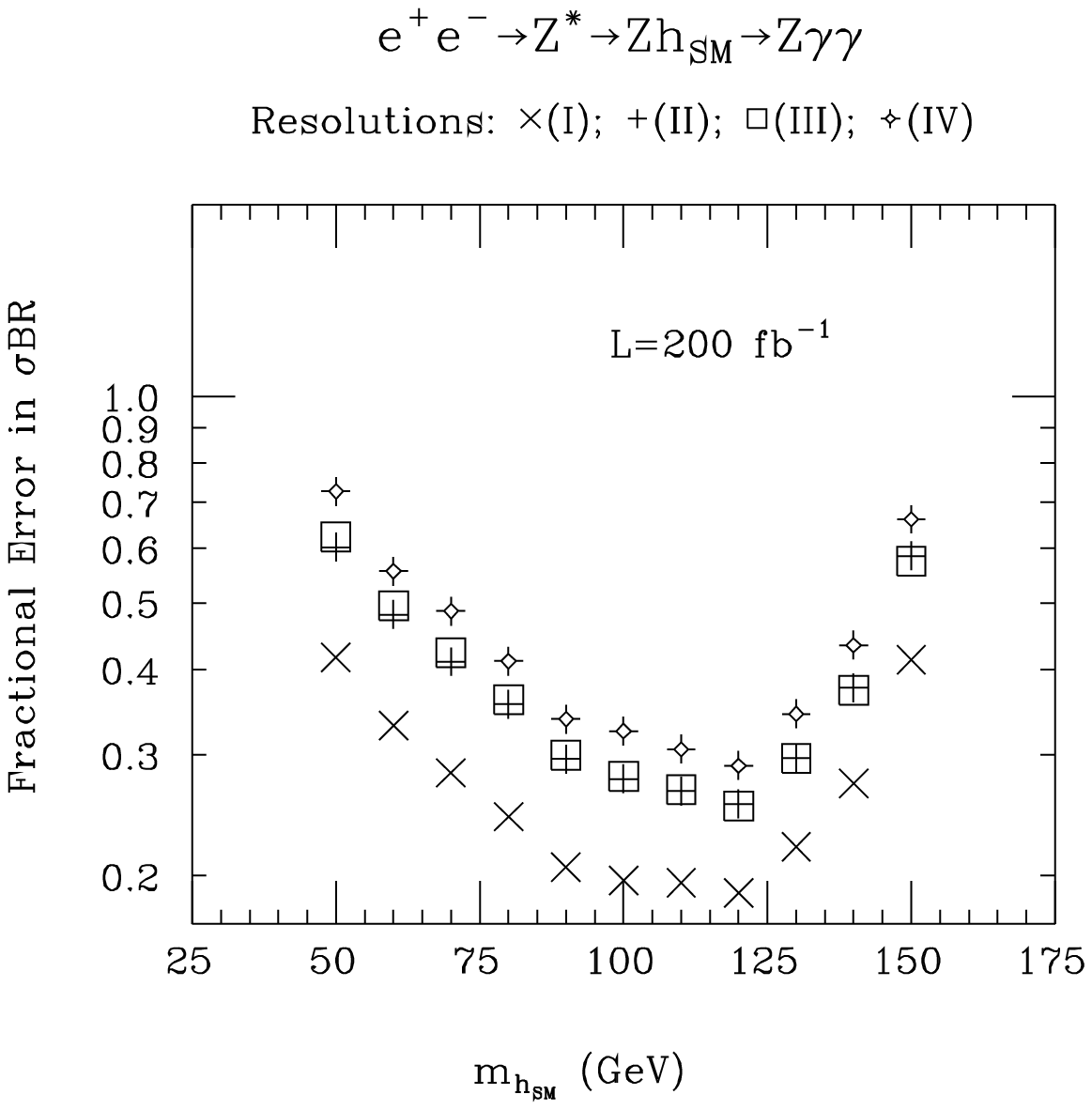,width=3.5in}}
%\bigskip
\end{center}
%\resizebox{!}{11cm}{%
%\includegraphics{3144A116.eps}}
%\end{center}
\caption{The fractional error in the measurement of
$\sigma(Z\hsm)\br(\hsm\to\gam\gam)$ 
 as a function of $\mhsm$.}
\label{fig3}
\end{figure}

\begin{figure}[htb]
\leavevmode
\begin{center}
\centerline{\psfig{file=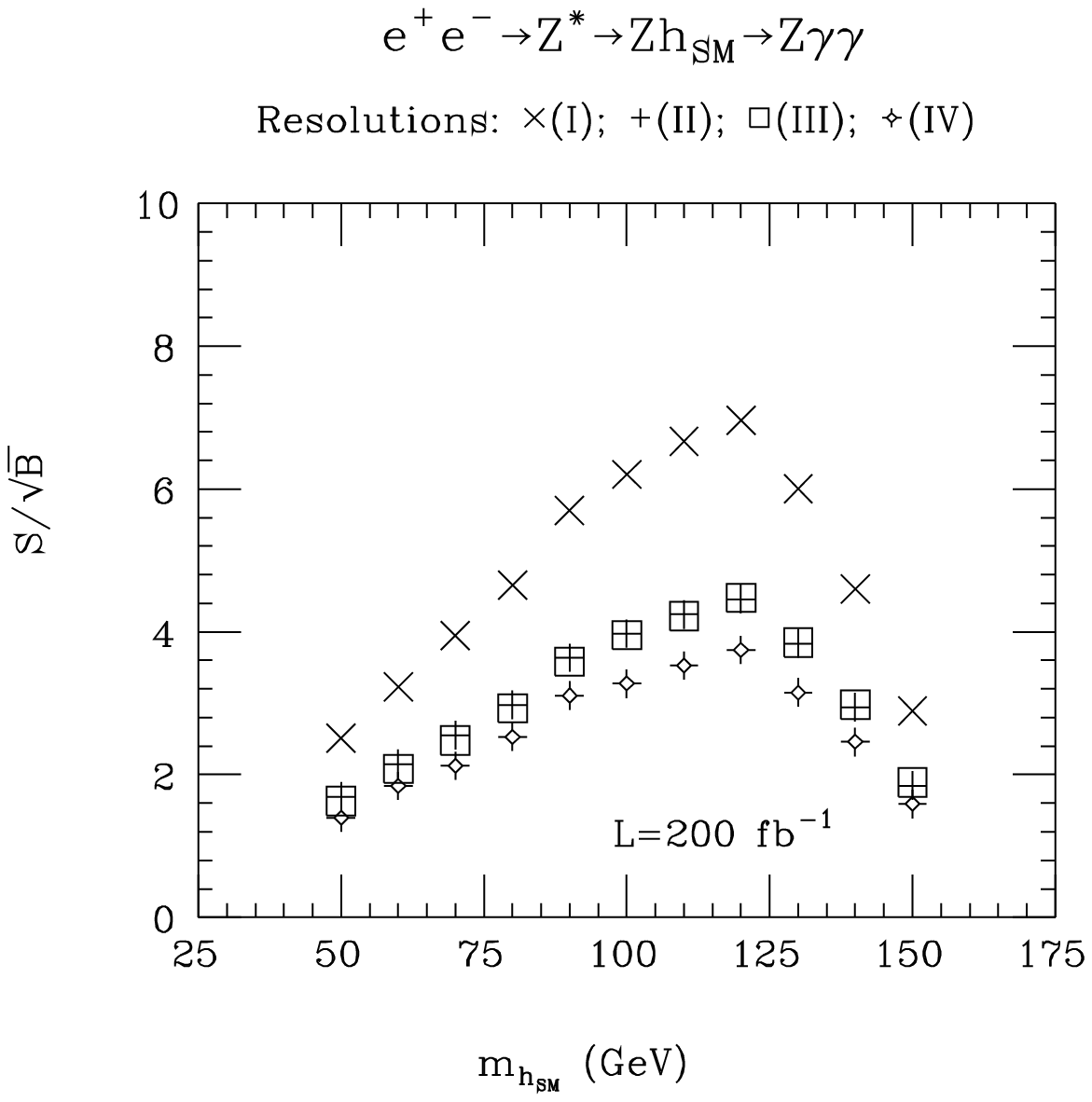,width=3.5in}}
%\bigskip
\end{center}
%\resizebox{!}{11cm}{%
%\includegraphics{3144A116.eps}}
%\end{center}
\caption{Results for $S/\protect\sqrt B$ in the $Z\hsm$ production
mode at $L=200\fbi$ as a function of $\mhsm$.}
\label{fig4}
\end{figure}

Figures~\ref{fig1}-\ref{fig4} display plots of the statistical error,
$\sqrt{S+B}/S$, and the statistical significance,
$S/ \sqrt B$, as functions of $\mhsm$ for both $\hsm\to\gam\gam$ 
measurement modes. The following observations are useful:
\begin{itemize}
\item
Figures~\ref{fig1} and \ref{fig3} reveal that in
resolution cases II-IV smaller errors
are obtained in the $Z\hsm$ measurement mode for a Higgs
mass between $50 \gev$ and $120 \gev$, whereas 
the $\wp\wm$ measurement mode yields 
smaller errors for $130 \lsim\mhsm\lsim 150\gev$.
In resolution case I, the $Z\hsm$ mode error is 
smaller for masses up to $130\gev$.
\item
The absolute minimal statistical error (as obtained
if we set $B=0$ and choose $\dmgamgam$ large enough
to accept the entire Higgs signal)
for $50\gev\lsim \mhsm\lsim 150\gev$ is:
\begin{description}
\item $\pm 8\%$ to $\pm 15\%$ in the $Z\hsm$ measurement mode; and
\item $\pm 15\%$ to $\pm 30\%$ in the $\wp\wm$ measurement mode.
\end{description}
These numbers indicate the extent to which the accuracy is limited
simply as a result of the very small event rates 
in the $\hsm \to \gam\gam$ decay mode. The smaller error
possible in the $B=0$ limit in the $Z\hsm$ measurement mode is 
a result of the larger $S$ values that can be achieved
by running at the optimal $\rts$.
\item
The smallest errors are obtained in the $90\gev\lsim \mhsm\lsim 130\gev$.
\footnote{This is the mass region predicted by the MSSM
for the light Higgs, $\hl$.} In this region the statistical
errors (including the computed background)
for the best calorimeter resolution case (case I) are:
\begin{description}
\item $\pm 19\%$ to $\pm 22\%$ in the $Z\hsm$ measurement mode;
\item $\pm 22\%$ to $\pm 32\%$ in the $\wp\wm$ measurement mode.
\end{description}
For the worst resolution case (case IV) the errors are:
\begin{description}
\item $\pm 29\%$ to $\pm 35\%$ in the $Z\hsm$ measurement mode;
\item $\pm 26\%$ to $\pm 41\%$ in the $\wp\wm$ measurement mode.
\end{description}
\item
Thus, if the detector does not have good 
electromagnetic calorimeter resolution,
then the $\wp\wm$--fusion measurement mode is quite competitive with,
and in some mass regions superior to,
the $Z\hsm$ measurement mode. However, if the smallest
possible errors are the goal, excellent resolution is required
and one must use the $Z\hsm$ measurement mode techniques
if $\mhsm\lsim 130\gev$.
The reasons behind these results are simple:
\begin{itemize}
\item
$S/B$ tends to be substantial in the $\wp\wm$--fusion mode, implying
relatively modest sensitivity to resolution, but $S$ itself
is limited (as noted earlier) 
so that even a $B=0$ measurement would not have a small error.
\item $S$ is larger in the $Z\hsm$ measurement technique
(at the optimal $\rts$ for the given $\mhsm$ value) but $B$ can
only be made small enough for a big gain in $\sqrt{S+B}/S$
if the mass interval accepted can be kept small.
\end{itemize}
\item
The plots also reveal that in the lower mass region, $50 \lsim\mhsm\lsim
80\gev$, the error in the $\sigma\br(\hsm\to\gam\gam)$
measurement would be substantially lower in the $Z\hsm$ mode,
whereas in the upper mass region of $140\lsim\mhsm\lsim 150\gev$
the errors are smaller in the $\wp\wm$ measurement mode, especially
if the resolution is not as excellent as assumed in case I.
\end{itemize}

Although observation of a clear Higgs signal
in the $\gam\gam$ invariant mass distribution
is not an absolute requirement (given that we will have observed the $\hsm$
in other channels and will have determined its mass very accurately)
it would be helpful in case there are significant systematics in
measuring the $\gam\gam$ invariant mass.  It is vital to be
certain that $\dmgamgam$ is centered on the mass region where
the Higgs signal is present. 
Figures~\ref{fig2} and \ref{fig4} show plots of the statistical significance,
$S/ \sqrt B$ {\it vs.} $\mhsm$. They show that the
mass regions for which $\geq 3\sigma$ measurements 
can be made depend significantly upon resolution.
\begin{itemize}
\item
If excellent resolution (case I) is available then $S/\sqrt B\geq 3$ is
achieved for $60\gev\lsim \mhsm\lsim 150\gev$ in both the $Z\hsm$
and $\wp\wm$--fusion measurement modes.
\item 
If the resolution is poor (case IV) then $S/\sqrt B\geq 3$
is achieved for $90\gev\lsim \mhsm\lsim 130\gev$
in the $Z\hsm$ measurement mode and for
$100\gev\lsim \mhsm\lsim 150\gev$ in the $\wp\wm$--fusion mode.
\end{itemize}

We end this section by noting that the error in the determination
of $\br(\hsm\to\gam\gam)$ is not precisely the same as
the error in the $\sigma\br(\hsm\to\gam\gam)$ measurement.
In the $\wp\wm$--fusion mode, Eq.~(\ref{stepiii})
shows that errors in both $\br(\hsm\to b\anti b)$ and
$\sigma(\nu_e\bar\nu_e\hsm)\br(\hsm \to b\anti b)$ enter into
the $\br(\hsm\to\gam\gam)$ error.  
The error in $\br(\hsm\to b\anti b)$ will be about $\pm8\%-\pm10\%$.
The error in $\sigma(\nu_e\bar\nu_e\hsm)\br(\hsm \to b\anti b)$
will probably be about $\pm 5\%-\pm 7\%$.  These errors must
be added in quadrature with the
$\sigma(\nu_e\bar\nu_e\hsm)\br(\hsm\to \gam\gam)$ error.
In the $Z\hsm$ measurement mode, $\br(\hsm\to\gam\gam)$ is computed
as $\sigma(Z\hsm)\br(\hsm\to\gam\gam)/\sigma(Z\hsm)$.  The $\sim\pm7\%$
error in $\sigma(Z\hsm)$ must
be added in quadrature with the $\sigma(Z\hsm)\br(\hsm\to\gam\gam)$ error.
However, since the $\sigma\br(\hsm\to\gam\gam)$ errors
in both the $\wp\wm$-fusion and $Z\hsm$ measurement modes
are always $\gsim\pm 20\%$, quadrature additions of the magnitude
summarized above will not be very significant.  For example,
for a $\sigma\br$ measurement of $\pm20\%$ the quadrature
additions would imply about $\pm 21\%$ ($\pm22\%$) 
errors for $\br(\hsm\to\gam\gam)$ using the
$Z\hsm$ ($\wp\wm$--fusion) measurement mode procedures.

\section{Conclusions}

We have studied the prospects for 
measuring $\sigma\br(\h\to\gam\gam)$ for a SM-like Higgs boson
at the NLC. The measurement will be challenging but of great importance.
We have compared results for two different production/measurement modes:
$\wp\wm$--fusion and $Z\hsm$. In the mass
range of $90 \gev$ to $130\gev$ where $\br(\hsm\to\gam\gam)$ is largest
(a mass range that 
is also highly preferred for the light SM-like $\hl$ of the MSSM)
the smallest errors in the measurement of $\sigma\br(\hsm\to\gam\gam)$
that can be achieved with an excellent CMS-style calorimeter 
(resolution case I)
are $\gsim\pm 20\%$ using the $Z\hsm$ measurement mode and $\gsim\pm 22\%$
using the $\wp\wm$--fusion measurement mode. For a calorimeter at the
optimistic end of current plans for the NLC detector (cases II and III)
the errors range from
$\sim\pm 25\%$ to $\sim\pm 30\%$ for the $Z\hsm$ mode and from
$\sim\pm 26\%$ to $\sim\pm 41\%$ for the $\wp\wm$--fusion mode.
The $Z\hsm$ errors assume that the machine energy
is tuned to the ($\leq 300\gev$) $\rts$ value which maximizes 
the $Z\hsm$ event rate,
and that $L=200\fbi$ is accumulated there, whereas the $\wp\wm$--fusion
errors assume that $L=200\fbi$ is accumulated at $\rts=500\gev$.

The desirability of running in the $Z\hsm$ measurement mode can only be 
determined once the Higgs mass is known. To take full advantage of such 
running would require that the calorimeter be upgraded to
a resolution approaching the CMS level of resolution.
For resolution cases II or III and $\mhsm\sim 120\gev$,
the accuracy of the measurement would be $\sim\pm 26\%$
in the $\wp\wm$--fusion measurement mode and $\sim \pm 25\%$ in
the $Z\hsm$ measurement mode, and there would be little point
in running in the latter mode. For CMS resolution (case I),
these respective errors become $\sim\pm 22\%$ and $\sim\pm 19\%$,
a gain that is still somewhat marginal, especially given the fact that
current estimates \cite{poggioli} are that the error 
on the $\sigma\br(\hsm\to\gam\gam)$ measurement at the LHC
would be comparable, of order $\pm 22\%$ at $\mhsm=120\gev$,
so that statistics could be combined to give $\sim \pm 15\%$
for either NLC measurement mode. However, for smaller $\mhsm$ values
the LHC error will worsen significantly and the $Z\hsm$
measurement mode becomes increasingly superior to the $\wp\wm$--fusion
mode, especially if the calorimeter resolution is excellent.
For Higgs masses above $\mhsm\sim 120\gev$, little would
be gained by using the $Z\hsm$ measurement mode; for the highest
mass considered, $\mhsm=150\gev$, using the $Z\hsm$ mode
would be disadvantageous.

\bigskip
\noindent
{\bf Acknowledgements} We thank J. Brau, R. Van Kooten, 
L. Poggioli and P. Rowson for helpful conversations.

\begin{table}[hbt]
\caption[fake]{For resolution choices I, II, III and IV,
we tabulate $p_T^{\gam_1~{\rm min}}$, $p_{T}^{\gam_2~{\rm min}}$, 
$\ptmin$, and $\dmgamgam$ as a function of $\mh$ (GeV).}
\begin{center}
\begin{tabular}{ccccccccc}
\hline
\hline
$\mh$ & \mbox{\scriptsize $p_T^{\gam_1~{\rm min}}$} & 
\mbox{\scriptsize $p_{T}^{\gam_2~{\rm min}}$} & 
\mbox{\scriptsize $p_T^{\rm min}$} &
\mbox{\scriptsize $\dmgamgam^I$} &
\mbox{\scriptsize $\dmgamgam^{II}$} &
\mbox{\scriptsize $\dmgamgam^{III}$} &
\mbox{\scriptsize $\dmgamgam^{IV}$} \\

\hline
50  & 30 & 10 & 40 & 0.7 & 2.0 & 2.1 & 2.3 \\
60  & 30 & 10 & 60 & 1.0 & 2.2 & 2.3 & 2.7 \\
70  & 30 & 20 & 65 & 1.1 & 2.4 & 2.7 & 3.2 \\
80  & 30 & 20 & 70 & 1.3 & 2.7 & 2.7 & 3.6  \\
90  & 30 & 20 & 70 & 1.4 & 3.1 & 3.1 & 4.1 \\
100 & 40 & 20 & 75 & 1.8 & 3.4 & 3.4 & 4.5 \\
110 & 40 & 20 & 85 & 2.0 & 4.2 & 4.2 & 5.0 \\
120 & 40 & 20 & 95 & 2.2 & 4.6 & 4.6 & 5.4 \\
130 & 50 & 20 & 100 & 2.3 & 4.9 & 4.7 & 5.9 \\
140 & 50 & 30 & 110 & 2.5 & 5.3 & 4.8 & 6.3 \\
150 & 50 & 30 & 120 & 2.4 & 5.4 & 5.1 & 6.8 \\
\hline
\hline
\end{tabular}
\end{center}
\label{tablei}
\end{table}

\begin{table}[hbt]
\caption[fake]{For resolution choices I, II, III and IV,
we tabulate $S$ and $B$ as a function of $\mhsm$ (GeV) for
$L=200\fbi$.}
\begin{center}
\begin{tabular}{ccccccccc}
\hline
\hline
$\mhsm$ & $S_I$ & $B_I$ & $S_{II}$ & $B_{II}$ & $S_{III}$ & $B_{III}$ 
& $S_{IV}$ & $B_{IV}$ \\

\hline
50 & 5.1 & 24 & 5.8 & 68 & 5.8 & 72 & 5.2 & 76 \\
60 & 8.1 & 24 & 8.0 & 54 & 8.1 & 56 & 7.6 & 66\\
70 & 8.8 & 12 & 8.4 & 27 & 8.7 & 30 & 8.3 & 35\\
80 & 12 & 13  & 12 & 29 & 12 & 29 & 12 & 37\\
90 & 18 & 15 & 17 & 31 & 17 & 31 & 17 & 40\\
100 & 22 & 14 & 20 & 27 & 20 & 27 & 20 & 35\\
110 & 26 & 13 & 25 & 27 & 25 & 27 & 24 & 31\\
120 & 29 & 11 & 27 & 23 & 28 & 23 & 26 & 27\\
130 & 26 & 9.3 & 25 & 20 & 24 & 19 & 24 & 23\\
140 & 18 & 6.1 & 18 & 13 & 17 & 12 & 17 & 15\\
150 & 12 & 4.7 & 12 & 11 & 12 & 10 & 11 & 13\\
\hline
\hline
\end{tabular}
\end{center}
\label{tableii}
\end{table}

\end{document}